\documentclass{aa}

\usepackage{graphicx}
\usepackage{txfonts}

\bibpunct{(}{)}{;}{a}{}{,} 

\newcommand{\rev}[1]{#1}

\usepackage{hyperref}

\makeatletter
\renewcommand*\aa@pageof{, page \thepage{} of \pageref*{LastPage}}
\makeatother

\hypersetup{%
        colorlinks,
        breaklinks=true,
        citecolor=[rgb]{0,0.20,0.45},
        linkcolor=[rgb]{0,0.20,0.45},
        urlcolor=[rgb]{0,0.20,0.45},
}
\begin{document}

   \title{Adding colour to the Zernike wavefront sensor: Advantages of including multi-wavelength measurements for wavefront reconstruction}


   \author{M. Darcis \inst{1,2}\fnmsep\thanks{Corresponding author; m.darcis@sron.nl}
        \and 
        S. Y. Haffert \inst{2,3}
        \and
        V. Chambouleyron \inst{4}
        \and
        D. S. Doelman \inst{1,2}
        \and
        P. J. de Visser \inst{1}
        \and
        M. A. Kenworthy \inst{2}
        }

   \institute{SRON - Space Research Organisation Netherlands, Niels Bohrweg 4, 2333 CA Leiden, The Netherlands
   \and
    Leiden Observatory, Leiden University, PO Box 9513, 2300 RA Leiden, The Netherlands
    \and
    Steward Observatory, The University of Arizona, 933 North Cherry Avenue, Tuscon, Arizona, USA
    \and
    University of California Santa Cruz, 1156 High St, Santa Cruz, USA
    }

   \date{Received 9 May 2025; Accepted 18 July 2025}

 
  \abstract
   {To directly image Earth-like planets, contrast levels of $10^{-8}$ - $10^{-10}$ are required. The next generation of instruments will need wavefront control below the nanometer level to achieve these goals. The Zernike wavefront sensor (ZWFS) is a promising candidate thanks to its sensitivity, which reaches the fundamental quantum information limits. However, its highly non-linear response restricts its practical use case.}
   {We aim to demonstrate the improvement in robustness of the ZWFS by reconstructing the wavefront based on multi-wavelength measurements facilitated by technologies such as the microwave kinetic inductance detectors (MKIDs).}
   {We performed numerical simulations using an accelerated multi-wavelength gradient descent reconstruction algorithm. Three aspects are considered: dynamic range, photon noise sensitivity, and phase unwrapping. We examined both the scalar and vector ZWFS.}
   {Firstly, we find that using multiple wavelengths improves the dynamic range of the scalar ZWFS. However, for the vector ZWFS, its already extended range was not further increased. In addition, a multi-wavelength reconstruction allowed us to take advantage of a broader bandpass, which increases the number of available photons, making the reconstruction more robust to photon noise. Finally, multi-wavelength phase unwrapping enabled the measurement of large discontinuities such as petal errors with a trade-off in noise performance.}
   {}


   \titlerunning{Advantages of including multi-wavelength measurements to the Zernike wavefront sensor}
   \maketitle
%
\section{Introduction} \label{section:introduction}

The next few decades will see a strong push towards the search for Earth-like planets in our cosmic neighbourhood.
To enable more insights to be drawn  with respect to the question of life in the universe, the atmospheres of these exoplanets need to be characterized. 
One of our most promising methods is to use high-contrast imaging instruments that resolve the planet from its host star \citep{snellen_detecting_2022}.
Because of its potential, a range of new instruments on the next generation of telescopes will be directed towards the direct imaging of exoplanets.
From the ground, there is the upcoming European Extremely Large Telescope, where most of the first-generation instruments will have some form of high-contrast imaging mode \citep{brandl_metis_2021,houlle_direct_2021, davies_micado_2016}.
A dedicated direct imaging instrument is being developed for the second generation called the Planetary Camera and Spectrograph \citep{kasper_pcs_2021}.
From space, NASA's Nancy Grace Roman Space Telescope will perform high-contrast observations and serve as a testbed for the more ambitious Habitable Worlds Observatory \citep{bailey_nancy_2023, national_academies_of_sciences_pathways_2023}.

To reach the desired contrast levels of the future ($10^{-8}$ for an exo-Earth around an M-dwarf and $10^{-10}$ for an exo-Earth around a Sun-like star), there are still significant technical challenges to be overcome.
A crucial part of this will be the wavefront sensing and control.
To reach the $10^{-10}$ contrast, the residual wavefront error can only be on the order of picometers \citep{ruane_vortex_2018}.
Therefore, a great deal of research is being dedicated to  building more sensitive and accurate wavefront sensors.
A highly promising technology is the Zernike wavefront sensor (ZWFS) because it is one of the most sensitive sensors and it even has the potential capacity to reach the fundamental quantum information limits for wavefront sensing \citep{bloemhof_phase_2003, haffert_arxiv_2023}.
This enables shorter integration times to reach a certain measurement performance, allowing for better contrast limits.
Furthermore, unlike the often used modulated pyramid wavefront sensor, it is inherently sensitive to aperture discontinuities such as segmentation and petaling, both of which significantly degrade the achieved contrast \citep{leboulleux_sensitivity_2018, bertrou-cantou_petalometry_2020}.
As a result, several groups are presently investigating its use in a high-contrast imaging setting, from measuring non-common path aberrations, as a second-stage atmospheric turbulence sensor, to segment co-phasing, and more \citep{ndiaye_calibration_2013, ndiaye_cascade_2024, salama_keck_2024}.

Essentially, a ZWFS is a self-referenced interferometer where part of the input wavefront is phase-shifted and which serves as the reference beam.
The phase shift is induced by a phase mask in a focal plane with a small circular dot centred on the PSF.
In the classical scalar ZWFS, the dot is a small dimple whose optical path difference (OPD) produces the phase shift. 

In general, the response of the ZWFS for monochromatic light can be expressed as

\begin{equation} \label{twowaveinterf}
    I_{out} = I_{in} + I_{ref} + 2I_{in}I_{ref}\cos(\phi_{in} - \phi_{ref}),
\end{equation}

where $I_{out}$ is the measured intensity, $I_{in}$ the intensity in the input pupil plane, and $I_{ref}$ and $\phi_{ref}$ are  the intensity and phase, respectively, of the self-created reference wave at the output pupil plane.
This equation highlights one of the main problems of the ZWFS, namely its dynamic range.
 To find $\phi_{in}$, the cosine needs to be inverted which is only possible over a $\pi$ interval. Furthermore, the presence of $\phi_{ref}$ in the cosine term makes the response asymmetric around $\phi_{in}$ and, thus, the range is not centred around 0.

 The full dynamic range on a $2\pi$ interval can be recovered by employing phase diversity, where multiple measurements are taken with different phase responses.
 The diversity can be introduced by adding a known phase difference to the input or by changing the phase shift of the ZWFS \citep{haffert_into_2024, wallace_phase-shifting_2011}.
 The latter can be implemented by using a vector ZWFS concept \citep{doelman_simultaneous_2019, chambouleyron_reconstruction_2024}.
 A vector ZWFS uses liquid crystals or metasurfaces to simultaneously induce a phase shift to one polarization state and the opposite phase shift to the orthogonal polarization state.

When trying to achieve better levels of precision, another limitation of the ZWFS is its chromaticity.
The ZWFS is a white light interferometer and can still work on a few ten per cents of bandwidth.
However, when going to wider bandwidths in order to capture more photons and lower the measurement noise, there are two main  chromatic effects that become important.
Firstly, the dimple of the scalar ZWFS provides a fixed OPD and, therefore, the induced phase-shift varies with wavelength, $\lambda$, as described by $(2\pi/\lambda) \cdot (n-1) \cdot d$. Here, $n$ and $d$ are the refractive index of the mask material and the depth of the dot, respectively. 
In contrast, the vector ZWFS does not suffer from this problem since its working principle is based on geometric phase which makes it possible to have the same phase shift for different wavelengths.
Secondly, due to the physically fixed diameter of the Zernike dot, the size of the dot relative to the PSF is different for each wavelength since the latter scales with $\lambda$.
This is an issue for both scalar and vector ZWFS.
In the past, certain methods have been proposed to achromatize the size of the dot, but this is not considered here \citep{bloemhof_phase-contrast_2004}.
As a result, since a change in phase shift and/or relative dot size changes the sensitivity of the ZWFS \citep{chambouleyron_modeling_2023, chambouleyron_variation_2021}, each wavelength experiences a different response.
These chromatic effects essentially \rev{reduce the fringe visibility and therefore the signal-to-noise ratio of the measurement. This, in turn, degrades} the reconstruction performance, limiting the usable bandwidth.

To further develop the ZWFS, multi-wavelength measurements offer a promising new avenue for improved performance.
\rev{In practice, this could already be implemented with existing techniques. If only a few wavelengths are desired,  it would be possible, for example, to implement a system that uses multiple dichroics to create different wavelength channels, as done in \citep{deo_spectral_2024, magniez_polychromatic_2024}. For a larger amount of measurements, another alternative can be to use one of the available integral field unit (IFU) technologies. Nevertheless, all these options will have their eventual limitations when pushed to the limits. For example, different wavelengths can experience different non-common path aberrations that complicate accurate wavefront reconstruction. In addition, truly broadband operation from the visible to infrared is prohibited by the use of semiconducting detectors. However, new technologies are on the horizon that can help mitigate these limitations.} 

\rev{One promising technology for use in multi-wavelength wavefront sensing is the microwave kinetic inductance detector (MKID), which can count individual photons and estimate their energy \citep{mazin_superconducting_2012, kouwenhoven_resolving_2023, magniez_mkid_2022, magniez_polychromatic_2024}.}
In this way, multi-wavelength measurements can be taken without the need for changing the optical architecture. 
Furthermore, compared to classical semiconducting detectors, MKIDs can also be rendered simultaneously sensitive all the way from the ultraviolet to the near-infrared. 
Resolving powers of R $\approx 50$ have been demonstrated in this wavelength range and will only improve with further research \citep{visser_phonon-trapping_2021}.
Moreover, the measurements will also benefit from the zero read noise due to the photon counting capability and the very low level of dark counts \citep{swimmer_characterizing_2023}.
In addition, arrays containing thousands pixels have already been implemented which is sufficient for wavefront sensing purposes \citep{walter_mkid_2020}.

In this paper, we examine the benefits of using multi-wavelength measurements in combination with a classic scalar ZWFS or a vector ZWFS.
In the limit of achromatic \rev{OPD} errors, three potential improvements are identified and investigated.
Firstly, different wavelengths experience a different phase for the same OPD and therefore offer a form of phase diversity that can be exploited to improve the dynamic range.
Secondly, wavelength information has the potential to broaden the bandpass and increase the robustness against photon noise.
Classically, a narrow wavelength range has been used since the response of the ZWFS is chromatic but when measurements at different wavelengths are separately available, these can be combined to effectively use the information of more photons to reconstruct the wavefront.
Thirdly, multi-wavelength measurements can be used to extend the dynamic range well beyond the $2\pi$ interval by enabling phase unwrapping, as is done in more classical interferometry \citep{cheng_two-wavelength_1984}.

 The work is outlined as follows.
 Section \ref{theory} starts with an overview of the non-linear reconstruction method used here, which also covers how multi-wavelength measurements can be included.
 This section also outlines the principle of two-wavelength phase unwrapping.
Next, Section \ref{performance} investigates the performance of a multi-wavelength ZWFS, describing the numerical simulations used to examine the dynamic range, the photon noise sensitivity, and the capabilities of two-wavelength phase unwrapping.
Our overall results and conclusions are given in Section \ref{dandc}.

\section{Multi-wavelength reconstruction} \label{theory}

\subsection{Accelerated gradient descent} \label{reconstructor}

For the monochromatic case, there are several ways to retrieve $\phi_{in}$ of the wavefront from Equation \eqref{twowaveinterf}.
The methods generally differ in the manner in which the reference field is modelled and how the cosine is inverted.
For the reference field, the simplest approach is to model it assuming a flat wavefront \citep{ndiaye_calibration_2013}.
More sophisticated methods iteratively update the reference field estimate \citep{doelman_simultaneous_2019, haffert_into_2024, chambouleyron_reconstruction_2024}.
Next, the cosine can be inverted as is or can be simplified by employing for example a Taylor series \citep{ndiaye_calibration_2013}.
Additionally, as mentioned before, phase diversity can be employed to overcome the limited invertibility range of the cosine.

One of the limitations of the  aforementioned methods is that they are defined for monochromatic light and do not translate well to a multi-wavelength scenario.
There exist methods that build a data-driven model using for example radial basis functions or neural networks that can incorporate multiple wavelengths \citep{lin_nonlinear_2024, allan_deep_2020}.
However, this work uses a different approach.
We made use of a recently proposed accelerated gradient descent method to solve the wavefront estimation problem that is also  able to incorporate multi-wavelength measurements in a relatively easy way (Haffert et al. in prep).
Here, the measured output is directly compared to the output of a model and an optimizer is used to find the parameters of the model that best describe the measurements.
Compared to the data-driven chromatic reconstructors such as neural networks, the model-based gradient descent method offers a more explainable framework, which makes it possible to better investigate the effects of employing multi-wavelength measurements.

The  parameters estimated in this work are the modal coefficients describing the OPD of the input wavefront.
Any desired mode basis can be used for the reconstruction in principle.
The cost function describing the difference between measurement and model is the maximum likelihood estimator and for the scalar ZWFS with multi-wavelength measurements can be described by

\begin{equation} \label{eq:costscalar}
   J(\theta) = \sum_{i}J_{i}(\theta) = \sum_{i} \frac{1}{N_{phot_i}^2} ||I_{out_i}^{measured}(\theta) - I_{out_i}^{modelled}(\theta)||_{2}^2,
\end{equation}

where $\theta$ are the OPD parameters and $||\cdot||_{2}$ is the two-norm.
Each wavelength has its own cost function, which are summed to produce the overall cost. Also,
$I_{out_i}^{measured}$, $I_{out_i}^{modelled}$, and $N_{phot_i}$ are  the measured output intensity,  modelled output intensity, and  total number of photons at a wavelength, $i$, respectively.
The factor $1/N_{phot,i}^2$ is a normalization factor to enable to work at various input levels.
We could further tweak the cost function, for example, by using a weighting system to control the contribution of each wavelength, but this type of cost function engineering is not considered here.

For a vector ZWFS, the cost function changes slightly because there are two pupil images per wavelength, one for each considered polarization state, and therefore can be described as 

\begin{equation} \label{eq:costvector}
   J(\theta) = \sum_{i}\sum_{j=0,1}J_{i,j}(\theta) = \sum_{i}\sum_{j=0,1} \frac{1}{N_{phot_{i,j}}^2} ||I_{out_{i,j}}^{measured}(\theta) - I_{out_{i,j}}^{modelled}(\theta)||_{2}^2.
\end{equation}

\begin{figure*} 
    \centering
    \includegraphics[width=\linewidth]{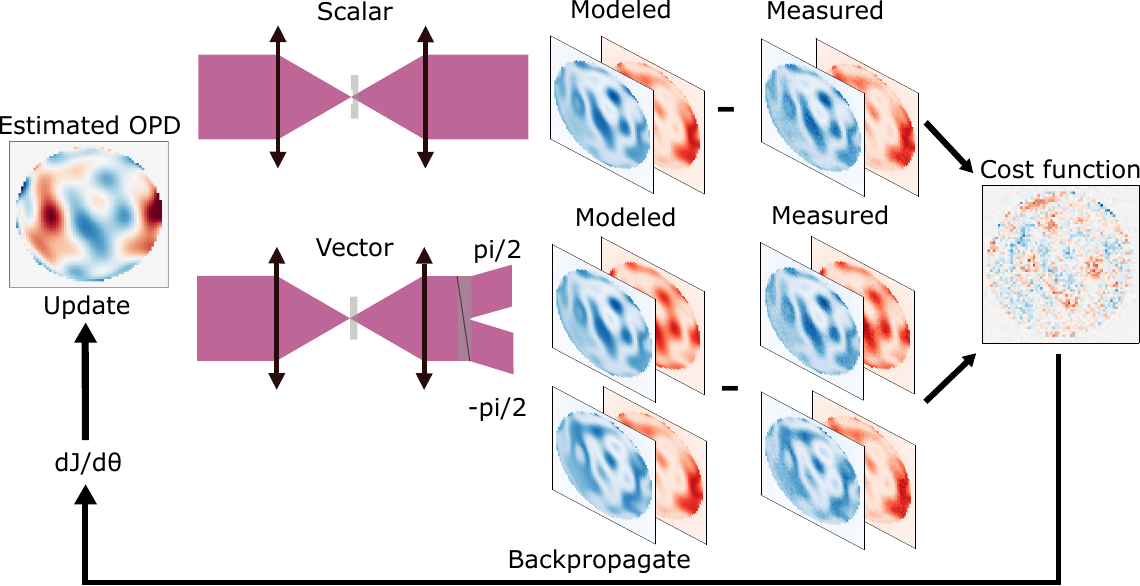}
    \caption{Gradient descent based wavefront reconstruction using multiple wavelength measurements for a scalar and a vector ZWFS.
    The input wavefront is described by a set of modal coefficients, $\theta$.
    The current estimate is propagated through a model of the considered ZWFS which produces the expect output intensity images at the considered wavelengths.
    These are compared to the measured intensity images and the similarity is captured in a cost function $J(\theta)$.
    The gradients of this cost function with respect to $\theta$ are calculated by a back-propagation process and these are used to update the current estimate by using an optimizer such as the Newton-CG.}
    \label{fig:overview}
\end{figure*}

A powerful way to minimize Equation \eqref{eq:costscalar} or \eqref{eq:costvector} is to analytically calculate the gradients using back propagation for an optical system as discussed in \citet{jurling_applications_2014}.
Using the assumption that each wavelength experiences the same OPD at the input, the gradients are calculated with respect to each cost function $J_{i,j}(\theta)$ separately and then averaged to produce the overall $dJ/d\theta$.
The gradients are then given to a solver which significantly speeds up the process, making it suitable for the real-time demands of extreme adaptive optics applications.
In this work, the Newton-CG method of SciPy is employed as the optimizer \rev{\citep{nocedal_large-scale_2006}}. In summary, Figure \ref{fig:overview} shows the overview of the reconstruction method for the scalar and the vector ZWFS setting.

\subsection{Phase unwrapping}

The interferometric nature of a ZWFS limits the dynamic range to a $2\pi$ interval because outside this range, the phase wraps and the measurement becomes ambiguous.
To break this ambiguity and unlock a significantly larger range, there are many algorithms that can be used to try and recover the true phase from the wrapped measurement based on spatial correlations \citep{schofield_fast_2003, herraez_fast_2002}.
These methods are not suitable for the next generation of segmented telescopes, where gaps between mirror segments destroy the spatial correlations.
Thick spiders holding a secondary mirror likewise cause discontinuities in the wavefront that lead to petal errors.
Phase unwrapping on these segmented apertures can still be done by employing correlations between wavelengths rather than spatial correlation, where the wrapped phase is measured separately at multiple wavelengths and these are then combined to solve for the unwrapped OPD \citep{cheng_multiple-wavelength_1985}.
In addition to being able to handle segmented apertures, these methods are computationally less expensive and so are better suited for high-speed adaptive optics applications.

In this paper, only two-wavelength phase unwrapping will be considered \citep{costa_quantitative_2010}, where two wavelengths are combined to produce an equivalent wavelength\rev{, also known as the beat wavelength,} of:

\begin{equation}
   \Lambda=\frac{\lambda_0\lambda_1}{|\lambda_1 - \lambda_0|}.   
\end{equation}

Considering the wrapped phase measurements $\phi_0$, $\phi_1$ on the interval $[0, 2\pi]$ at $\lambda_0$ and $\lambda_1$ respectively, the unwrapped OPD on the interval $[0, \Lambda]$ can then be calculated via

\begin{equation} \label{eq:phaseunwrap}
   OPD = \frac{\Lambda}{2\pi}[(\phi_0 - \phi_1 )\  \text{mod}\ 2\pi].   
\end{equation}

Earlier works have demonstrated that this method, together with a sparse aperture mask, can reconstruct large petal errors \citep{deo_spectral_2024}.
It is nevertheless interesting to test the combination with a ZWFS because of its sensitivity and its inherent ability to sense segmentation and petal errors. 
\citet{vigan_-sky_2011} have shown that large segmentation errors can be measured in this manner but only by fitting a very specific model for segment OPD differences in order to retrieve the wrapped phases.
Here, a monochromatic version of the gradient descent method outlined in Section \ref{reconstructor} is used to estimate $\phi_0$ and $\phi_1$, which, in principle, can be easily adapted to different types of errors by choosing the mode basis that is being reconstructed.
This aspect is further explored in Section \ref{section:pu}.

\section{Multi-wavelength ZWFS performance simulations and analysis} \label{performance}

To investigate the performance of a multi-wavelength ZWFS,  we performed simulations using the HCIPy package \citep{por_high_2018}.

\subsection{Zernike mask design}  \label{z_mask}

We must first find a mask design has to be found that works with multi-wavelength measurements.
The mask of a ZWFS has two design parameters: the phase shift introduced by the dot and the size of the dot, but carrying out a full optimization of mask design and accompanying cost function is beyond the scope of this paper.
Instead, the parameters are selected based on a few principles. 

Regarding the phase shift, the classical $\pi/2$ radians is taken as the baseline since it has been shown to provide the maximum sensitivity under read noise in linear conditions \citep{chambouleyron_modeling_2023}.
Furthermore, it has been established that an idealized wavefront sensor that shifts the piston mode by $\pi/2$ radians is able to achieve the fundamental sensitivity \citep{chambouleyron_coronagraph-based_2024}.
\rev{In addition, it is known that when two ZWFS measurements are used with different phase shifts to obtain phase diversity, then the optimal diversity is achieved when the difference in phase shift equals $\pi \; mod \; 2\pi$ radians \citep{chambouleyron_reconstruction_2024}.}  

To choose the phase shift for a scalar ZWFS mask in a multi-wavelength setting, we need to take into account the wavelength dependence of the induced phase shift as discussed in Section \ref{section:introduction}.
Remembering that phase is a $2\pi$ modular quantity, a certain desired phase shift is only achieved at specific wavelengths.
These can be tuned by changing the physical depth of the dot, which then changes the scaling relationship of the induced phase.
The vector ZWFS mask is easier to design since it can apply the phase shift achromatically and the $\pi$ phase difference can be directly implemented by applying a $\pi/2$ and $-\pi/2$ phase shift to two different polarization states.

\rev{The main consideration for choosing the diameter of the dot is to have each wavelength have a similar sensitivity. Each measurement will then have a comparable cost function fraction and the gradients will be of similar magnitude. This makes the algorithm easier to use and understand, without going into cost function engineering. It has been shown that increasing the dot size with respect to the PSF increases the sensitivity of the ZWFS, at the cost of lowering the sensitivity of lower order modes \citep{chambouleyron_variation_2021}. Therefore, wavelengths are chosen not too far apart in order to have a sufficiently comparable sensitivity over the full range of modes. In practice, it was seen that keeping the dot size limited to $\sim$1-2$\lambda/D$ produced the best results.}

\subsection{Dynamic range} \label{dr}

A two-wavelength simulation was performed to investigate the dynamic range, considering the following mask parameters.
Firstly, an arbitrary design wavelength $\lambda_{0}$ of $600$ nm was selected.
For the scalar mask, a phase shift of $5\pi/2$ at $\lambda_{0}$ is taken.
To provide the $\pi$ phase shift difference, the wavelength corresponding to a phase shift of $3\pi/2$ was chosen as the second wavelength, $\lambda_{1}$, which in this case is $1000$ nm.
The dot size was set to $2 \lambda/D$ at $\lambda_{0}$.
The dot size at $\lambda_{1}$ was then $1.2\lambda/D$.
For the vector mask, the same wavelengths and dot size were taken, but a phase shift of $\pm\pi/2$ was induced at each wavelength.

\begin{figure*} 
    \centering
    \includegraphics[trim={0cm 0cm 0cm 0cm},clip,width=\linewidth]{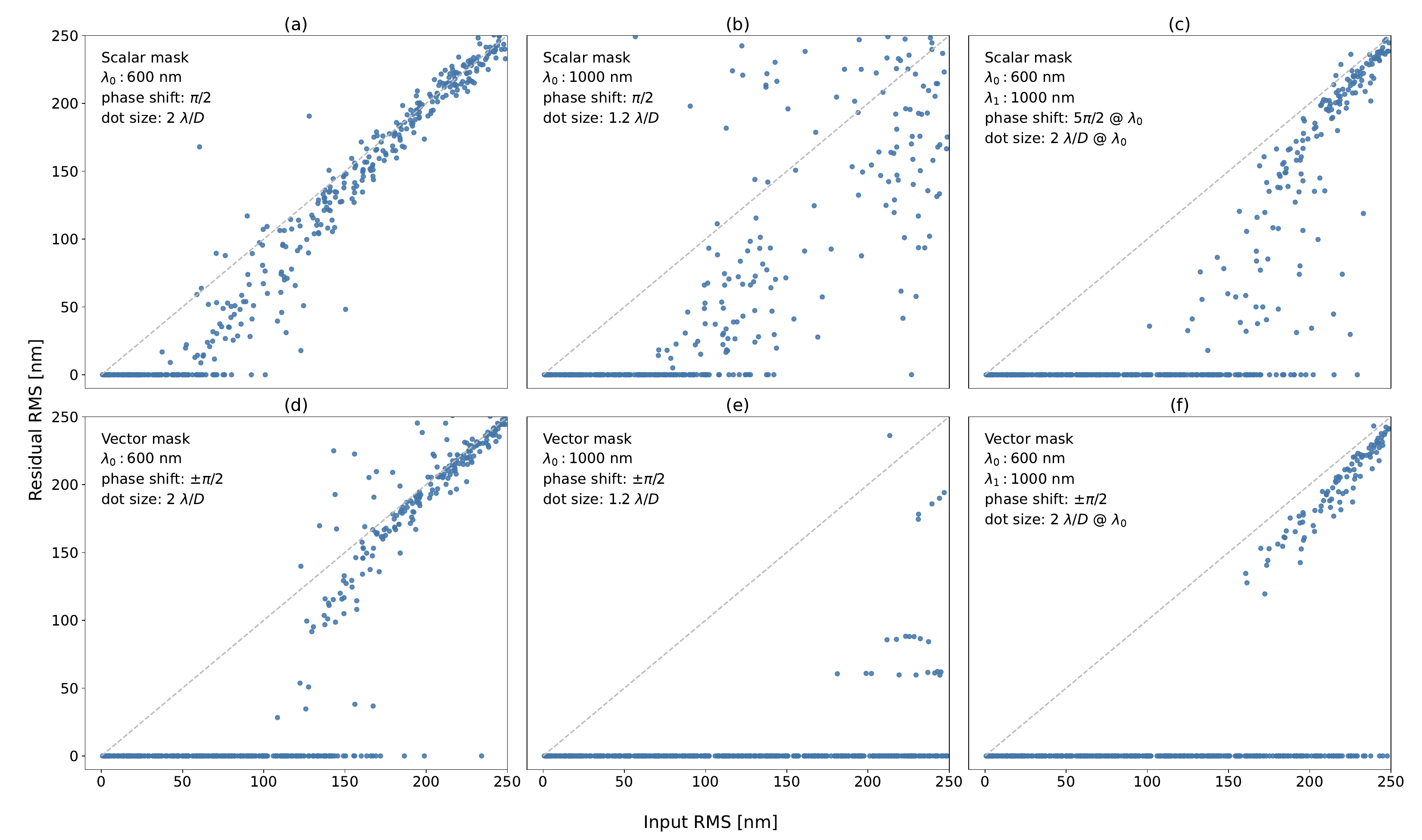}
    \caption{Dynamic range of the scalar and vector ZWFS using the gradient descent based reconstruction.
    For each configuration, 500 wavefronts are reconstructed consisting of 75 Zernike modes with a power law exponent between -1 and -3 and an RMS up to 250 nm.
    (a) and (b) Residual RMS for the scalar ZWFS using a monochromatic reconstruction for two different wavelengths (600, 1000 nm).
    (c) Residual RMS when employing the multi-wavelength reconstructor.
    (d), (e), and (f) Same results, but  for a vector ZWFS.}
    \label{fig:dynamicrange}
\end{figure*}

This allowed us to create a dataset of 500 OPD screens on a circular pupil of unit diameter.
For each screen, a power law error was generated on a $64\times64$ grid with a random exponent uniformly distributed between $-1$ and $-3$.
Then, 75 Zernike coefficients were calculated on the same $64\times64$ grid starting from defocus and serve as the description of the OPD error.
The RMS values were set to be uniformly distributed between $0$ and $250$ nm.
An example of a generated OPD screen can be seen in Figure \ref{fig:overview}.
Each OPD screen is initially reconstructed using a monochromatic version of the gradient descent method at $\lambda_{0}=$ 600 nm and $\lambda_{1}=$ 1000 nm separately and then reconstructed by combining the multi-wavelength measurements.
Figure \ref{fig:dynamicrange} compares the RMS of the input screens to the residual RMS errors left after subtracting the reconstruction for both the scalar and vector ZWFS.

Looking at the reconstruction result for the scalar ZWFS, the longer wavelength can successfully reconstruct larger OPD errors, compared to the shorter wavelength (as expected).
Nevertheless, when combining the two wavelengths, an even larger dynamic range can be achieved.
This shows that the algorithm is able to exploit the diversity provided by the different measurements.

For the vector ZWFS,  the longer wavelength once again provides a larger dynamic range.
However, combining the wavelengths no longer provides an improvement.
The diversity from having opposite phase shifts for different polarization states is sufficient to extend the dynamic range.
The additional diversity gained from having multiple wavelengths seems to be redundant and even makes the algorithm fail more often on larger OPD errors.

Looking closer at the input screens, we found that for both the multi-wavelength scalar and vector ZWFS, there are screens that show regions of phase wrapping at $\lambda_0$ and $\lambda_1$ that the algorithm was able to reconstruct (screens with roughly $>$175 nm RMS). 
The monochromatic gradient descent was also able to perform phase unwrapping in the vector ZWFS case.
The gradient descent method in itself is somewhat able to reconstruct phase wraps, even without the multi-wavelength measurements for the vector ZWFS. 
The hypothesis is that the dynamic range is increased by exploiting the mode mixing of the Zernike modes, where when exciting a single mode in the non-linear regime the signal starts to appear in other modes.
This capability is limited as the reconstructor starts to fail at larger RMS screens.
In the low Strehl regime, the interference between input and reference field becomes of too low contrast.

\begin{figure*} 
    \centering
    \includegraphics[trim={0.5cm 0.5cm 0.5cm 0},clip,width=\linewidth]{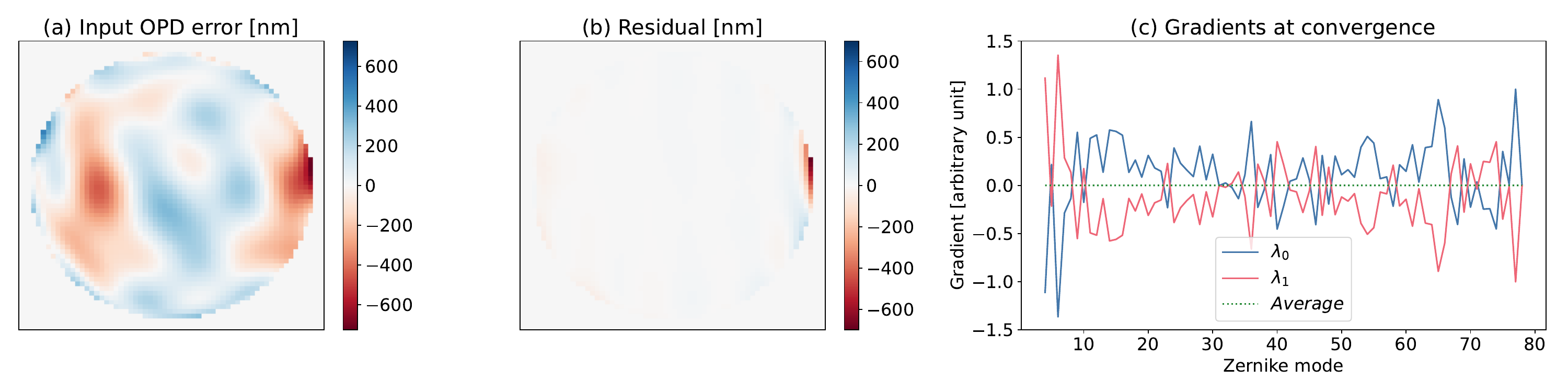}
    \caption{Example of the multi-wavelength gradient descent algorithm not finding the correct solution. The scalar ZWFS is used with wavelengths 600 and 1000 nm and the mask consists of a dot with $5\pi/2$ phase shift and 2 $\lambda/D$ dot diameter at 600 nm. (a) Input wavefront of 168 nm RMS. (b) Residual error after reconstruction of 38 nm RMS. (c) Calculated gradients at convergence at each wavelength and the overall gradient obtained by averaging.}
    \label{fig:failure}
\end{figure*}

An example of input and reconstruction residual is shown in Figure \ref{fig:failure} for the multi-wavelength scalar ZWFS for one of the OPD screens where it had been unsuccessful in finding the correct solution.
The algorithm successfully converged but is unable to fully reconstruct the edge.
Figure \ref{fig:failure} also plots the calculated gradients at the converged point.
It shows that the gradients of the different wavelength images cancel each other.
This makes the total gradient zero and appears to be a solution.
However, when noise is introduced the gradients will not perfectly cancel any longer and the interpretation will depend on the signal-to-noise ratio.

\subsection{Photon noise} \label{section:photonnoise}

Apart from the dynamic range, real systems also need to be robust with respect to  sources of noise, of which photon noise is the most fundamental.
Under only photon noise, the variance on the measurement of a specific phase mode $\phi_{i}$ under linear conditions is given by \citep{chambouleyron_modeling_2023}:

\begin{equation} \label{eq:photonnoise}
    \sigma^{2}_{\phi_{i}} = \frac{1}{s^{2}_{\gamma}(\phi_{i}) \cdot N_{phot}},   
\end{equation}

where $N_{phot}$ is the number of photons used for the measurement and $s_{\gamma}(\phi_{i})$ is the photon noise sensitivity, which is equal to the Fisher information and is derived to be $0 \leq s_{\gamma} \leq 2$ \citep{paterson_towards_2008}.

\begin{figure} 
    \centering
    \includegraphics[trim={0.7cm 0.2cm 0.7cm 0.5cm},clip,width=\columnwidth]{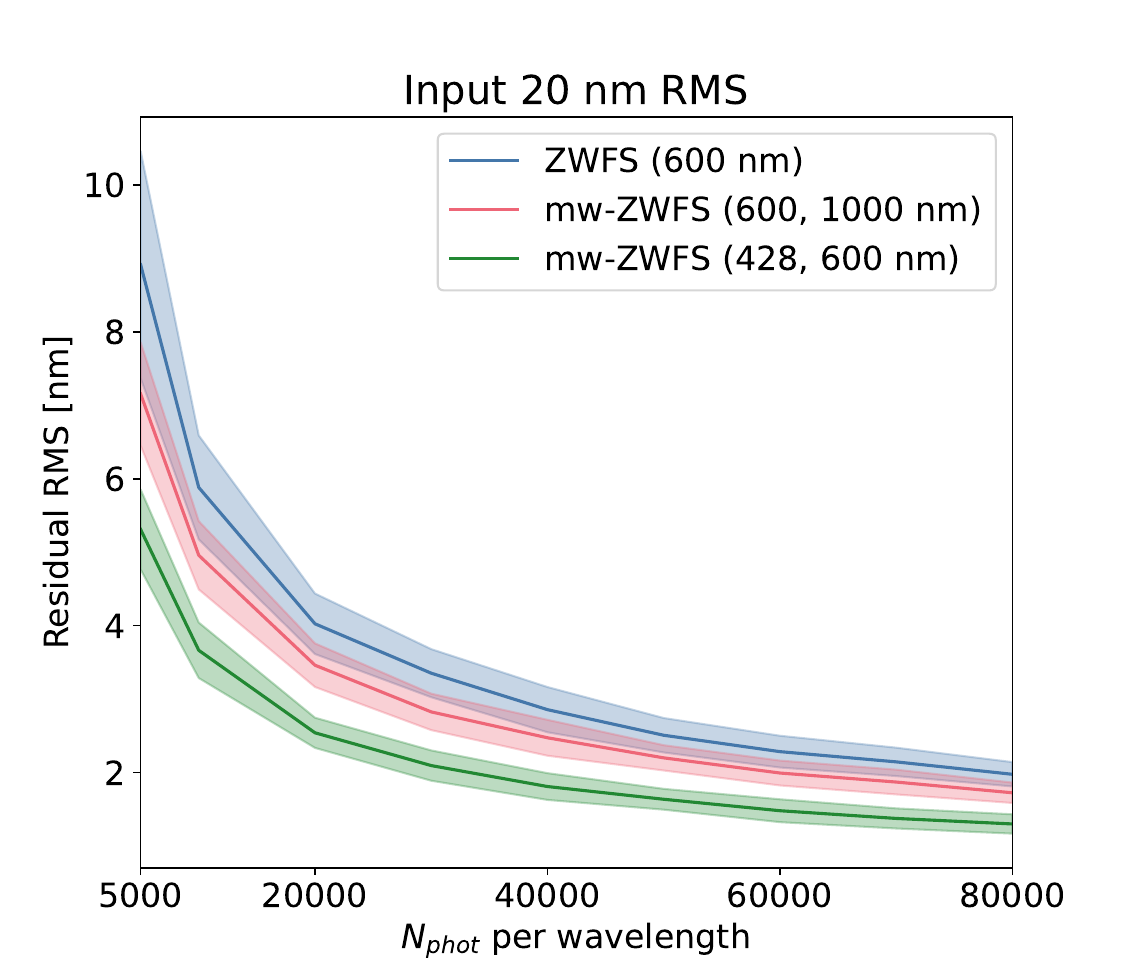}
    \caption{Reconstruction error at different photon levels for the monochromatic scalar ZWFS and the multi-wavelength scalar ZWFS using 600 nm and either 428 or 1000 nm. It is assumed that each wavelength has the same number of photons. 100 photon noise samples are reconstructed at each considered input level. \rev{Coloured areas correspond to $\pm1$ standard deviation.}}
    \label{fig:photonnoisesensitivity}
\end{figure}

By using multiple wavelengths for the measurement, the total number of photons is increased and according to Equation \eqref{eq:photonnoise}, this decreases the variance.
For the multi-wavelength reconstruction considered here OPD and not phase is the reconstructed quantity.
The variance of the OPD measurement is the variance of the phase times the factor $(\lambda/2\pi)^2$; in other words, shorter wavelengths have a larger phase for a fixed OPD and therefore produce a larger signal at the ZWFS output.
Not all photons, therefore, will contribute equally in increasing the signal-to-noise ratio.
To illustrate this point, Figure \ref{fig:photonnoisesensitivity} shows the reconstruction error under photon noise for the scalar ZWFS at 600 nm and two multi-wavelength scalar ZWFS configurations that have an additional wavelength at either 428 or 1000 nm, both of which experience a $3\pi/2$ mod $2\pi$ phase shift when using the same mask parameters as in Section \ref{dr}.
An OPD screen with an exponent of -2 and 20 nm RMS was taken as the input. Then,
100 photon noise samples were reconstructed at each considered input level.
Both the multi-wavelength configurations improve the reconstruction error, but by a different amount, even though they both have the same total number of photons. 

In a more generalized setting, we also need to take into account the fact that different wavelengths see a slightly different ZWFS configuration, which changes the sensitivity, $s_{\gamma}$.
An advantage of the vector ZWFS in this case is that it applies the same phase shift to the various wavelengths.
This has been shown before to improve the bandwidth over which a successful reconstruction is possible, thus allowing more photons to be part of the measurement \citep{doelman_simultaneous_2019}.
Conceptually, the signals at different wavelengths add up and under a flat spectrum the signals from the short and long wavelengths average out to produce a larger signal at the centre wavelength. 

\begin{figure} 
    \centering
    \includegraphics[trim={1cm 0.2cm 1cm 0.2cm},clip,width=\columnwidth]{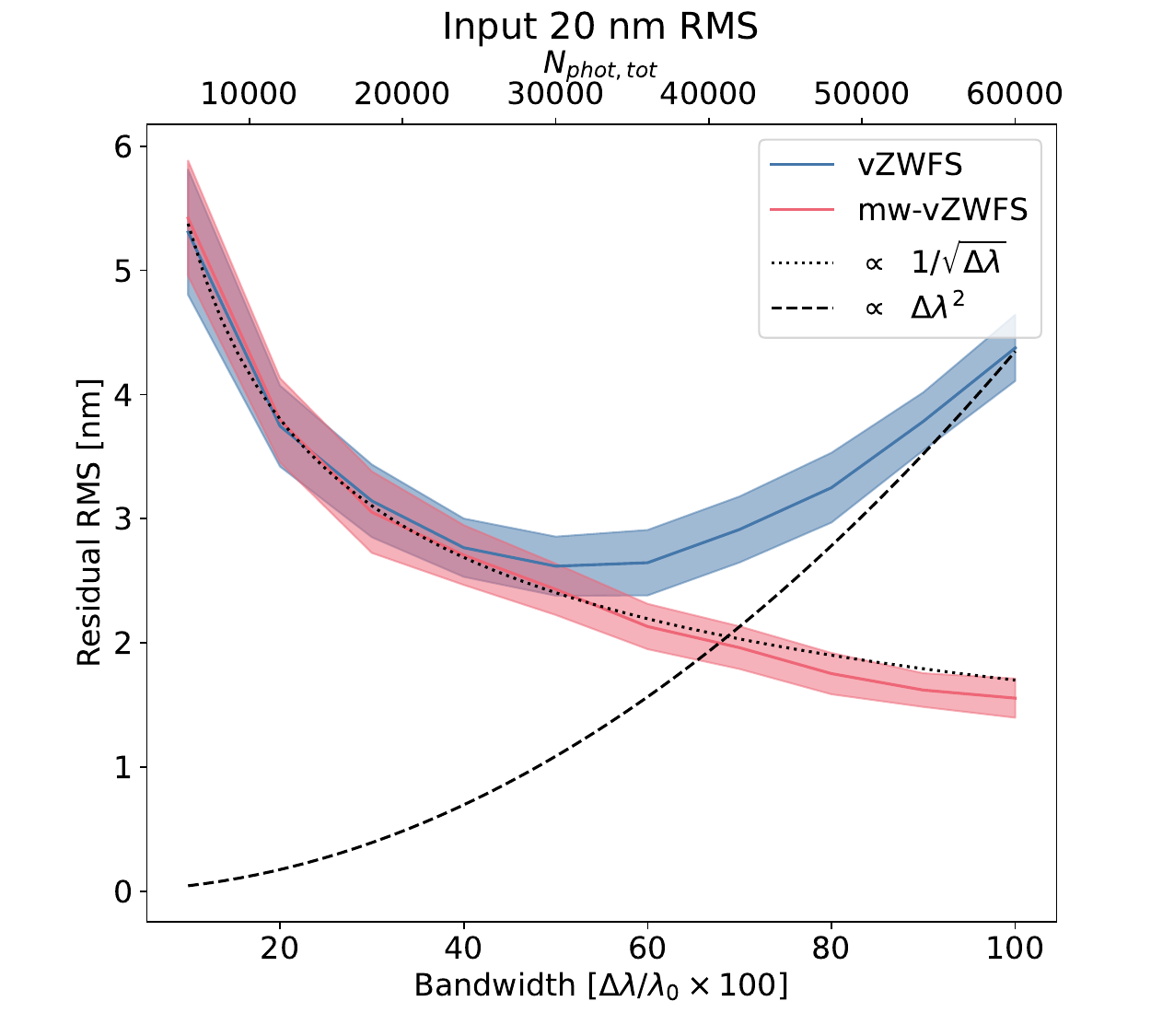}
    \caption{Reconstruction error at different bandwidth sizes using different reconstruction approaches for the vector ZWFS. The first method (vZWFS) does not separate the wavelengths and reconstructs at the central wavelength using the broadband image. The second method (mw-vZWFS) separates the bandwidth into ten wavelength bins and then uses the multi-wavelength algorithm. Then, 50 photon noise samples are taken at each bandwidth size. \rev{Coloured areas correspond to $\pm1$ standard deviation.}}
    \label{fig:bandwidth}
\end{figure}

To investigate whether multi-wavelength measurements can further broaden the useful bandwidth, the noise performance of a vector ZWFS with monochromatic reconstruction at the centre wavelength was compared to the scenario where the bandwidth is split up into multiple wavelength bins and the multi-wavelength algorithm was used for the reconstruction.
An OPD screen was generated with an exponent of -2 and 20 nm RMS.
A flat spectrum with a central wavelength of 600 nm was assumed for the simulations.
For each considered bandwidth size $\Delta\lambda$, the total number of photons was divided between ten equally spaced wavelength samples, which were then propagated through the vector ZWFS model.
A dot diameter of 2 $\lambda/D$ at 600 nm was taken.
For the standard vector ZWFS, the output signals were added together; whereas for the multi-wavelength version, they were considered separately in the reconstruction.
For each bandwidth size, 50 samples with a different photon noise realization were reconstructed and the performance is given in Figure \ref{fig:bandwidth}.
The reconstruction error is similar up to 50\%, after which the error for the classic vector ZWFS starts to go up again; whereas for the multi-wavelength version, the error continues to decrease.
This can be explained by considering the scaling of the error due to photon noise and due to broadband effects in function of bandwidth size.
For photon noise, the error scales according to $1/\sqrt{\Delta\lambda}$.
With some simplifying assumptions, it has been shown that the error induced by using a monochromatic reconstructor for broadband measurements scales proportionally to $\Delta\lambda^2$ \citep{haffert_into_2024}.
Increasing $\Delta\lambda$ reduces photon noise error but increases the chromatic errors.
Consequently, for the monochromatic reconstructor, there is an optimal $\Delta\lambda$ beyond which a broader bandwidth only increases the error, as seen in Figure \ref{fig:bandwidth}.
This point where broadband effects take over depends on the level and shape of the input spectrum.
However, when using multi-wavelength measurements the chromaticity can be taken into account and the error continues to scale in a way that resembles the photon noise for larger $\Delta\lambda$, increasing the usable bandwidth.
The improvement could be even greater in more realistic situations where the spectral type of the host star and chromaticity of the instrument are taken care of by the multi-wavelength reconstruction.

\subsection{Phase unwrapping} \label{section:pu}

To investigate the potential of multi-wavelength phase unwrapping, we performed simulations to examine its ability to reconstruct petal errors.
This application was chosen because the more common wavefront sensors such as the Shack-Hartmann and the modulated pyramid have difficulties in sensing these errors, which can lead to large petal jumps during operation.
Petal modes are fully orthogonal since they are defined on different parts of the aperture, so they do not show mode mixing. Thus, the gradient descent method on its own cannot perform any phase unwrapping, as described in Section \ref{dr}.

\begin{figure} 
    \centering
    \includegraphics[trim={0.07cm 0.2cm 0.5cm 0.3cm},clip,width=\columnwidth]{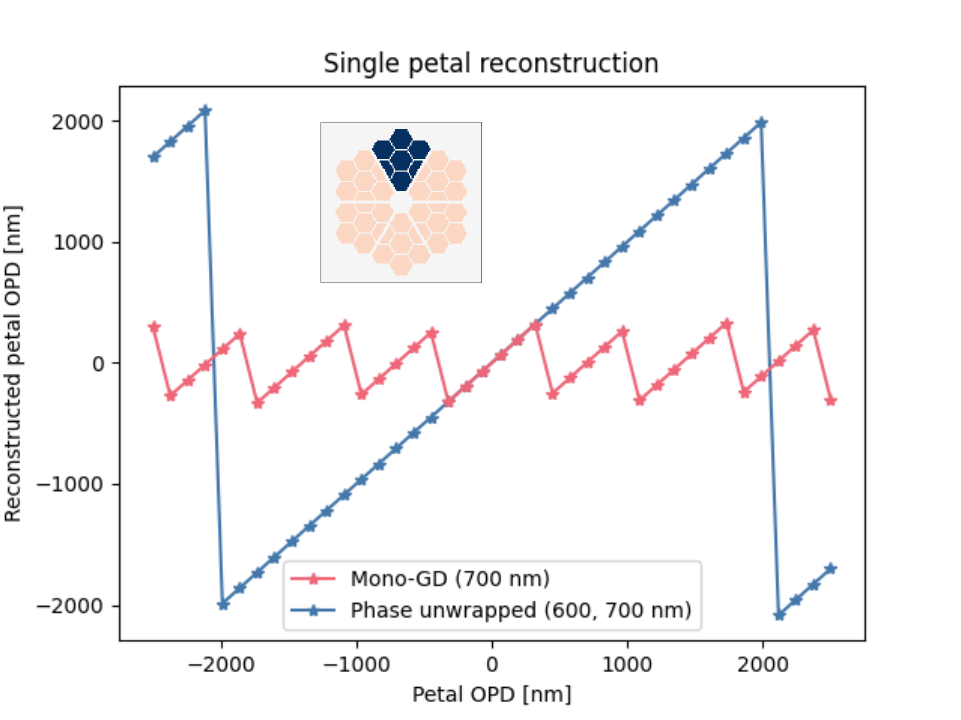}
    \caption{Reconstruction of a single petal mode using one wavelength at 700 nm (Mono-GD) and using two-wavelength phase unwrapping with 600 and 700 nm.}
    \label{fig:phaseunwraprange}
\end{figure}

The assumed pupil for the simulation has a general hexagonal segmented shape with six spiders and 36 segments.
An example of a displaced petal is shown in Figure \ref{fig:phaseunwraprange}.
Next, the vector ZWFS was used because it unlocks the $2\pi$ range at individual wavelengths, which is required by Equation \eqref{eq:phaseunwrap}.
The monochromatic gradient descent method was utilized to reconstruct the wrapped OPD of the six petal modes at the different wavelengths, which were then converted to phase to produce $\phi_0$ and $\phi_1$.
Afterwards, the unwrapped OPD error was reconstructed according to Equation \eqref{eq:phaseunwrap}.
The two wavelengths chosen for the simulation were 600 and 700 nm.
This offered an equivalent wavelength of 4.2 $\mu m$.

\begin{figure*} 
    \centering
    \includegraphics[trim={3cm 0.2cm 3cm 0.2cm},clip,width=2\columnwidth]{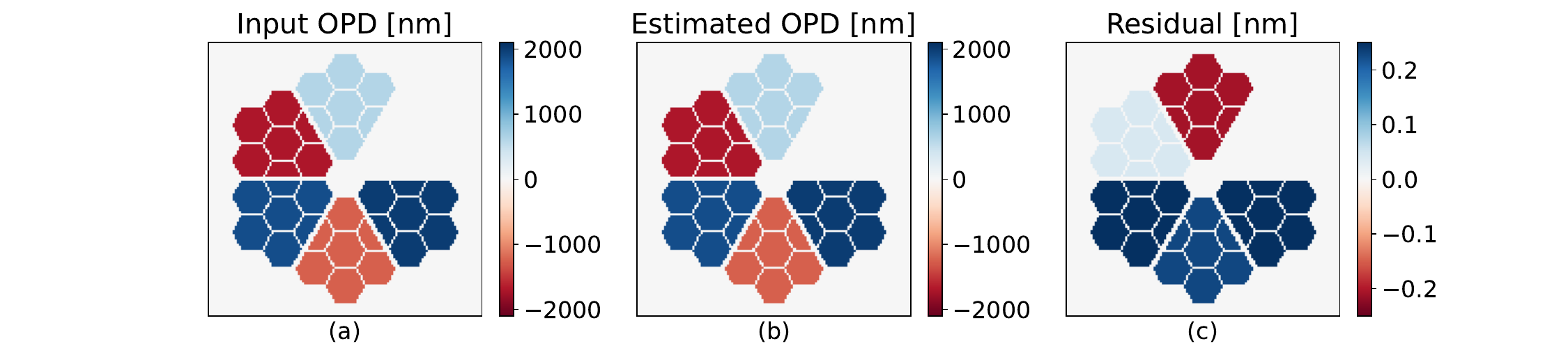}
    \caption{Reconstruction with multiple petal modes excited with a total RMS error of 1.4 $\mu$m. (a) the input wavefront. (b) the estimated wavefront. (c) the residual after subtracting the reconstruction from the input (different scale).}
    \label{fig:unwraprms}
\end{figure*}

Firstly, a single petal was displaced in 40 steps between $\pm$2.5 $\mu$m.
Figure \ref{fig:phaseunwraprange} shows the reconstructed petal OPD when using just a single wavelength of 700 nm and when using the two-wavelength phase unwrapping method.
As expected, the monochromatic reconstruction was limited to $\pm$350 nm, while phase unwrapping can reconstruct over the full $\pm$2.1 $\mu$m range.
To furthermore investigate the ability of handling also larger RMS errors, all petals except one (which is needed as a reference) were displaced by some amount.
Figure \ref{fig:unwraprms} shows the reconstruction results.
The gradient descent method can still find the correct wrapped phase maps also in a higher RMS regime.
This shows the potential of using multi-wavelengths for measuring large petal errors.

\begin{figure} 
    \centering
    \includegraphics[trim={0.5cm 0.5cm 0.5cm 0.5cm},clip,width=\columnwidth]{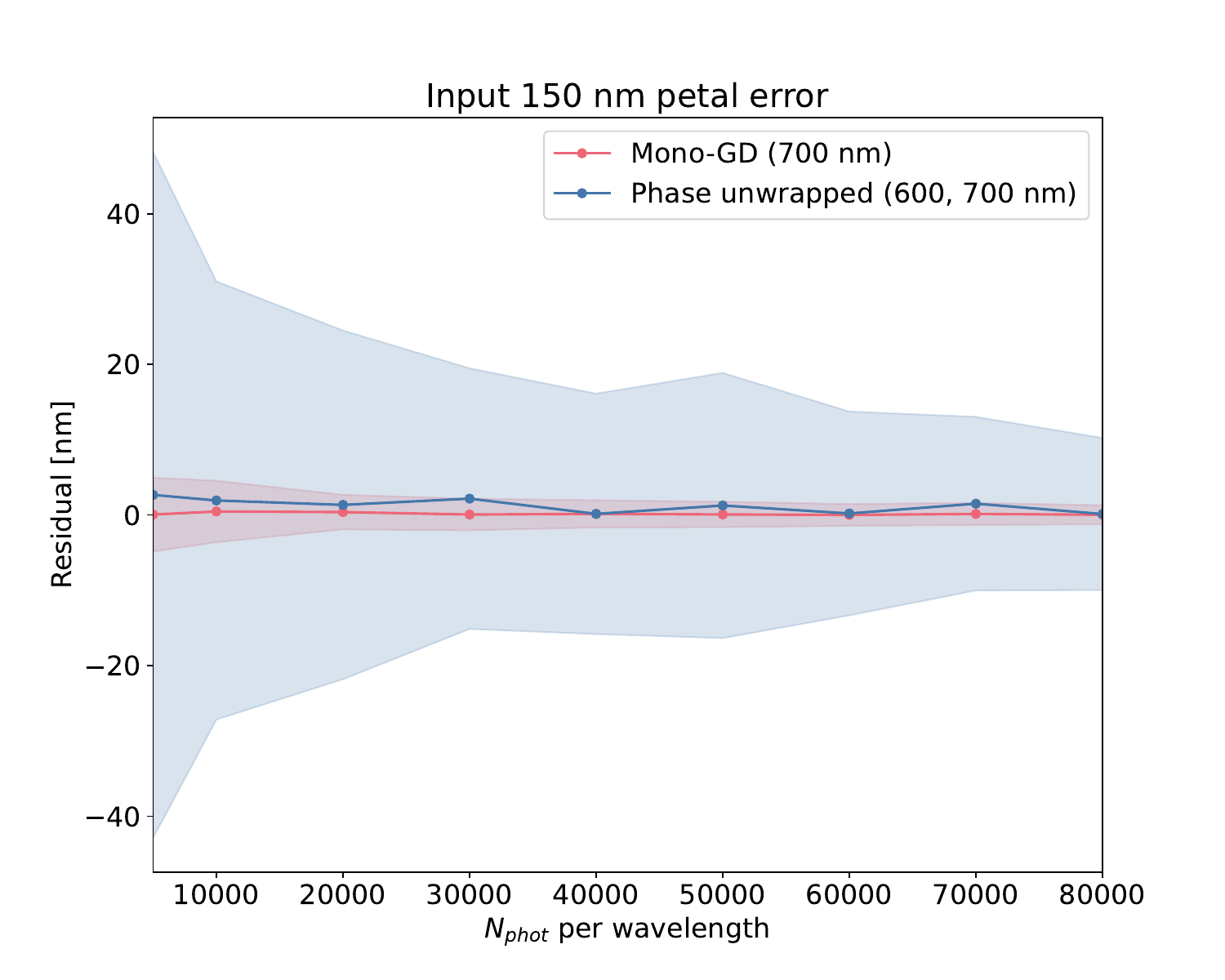}
    \caption{Reconstruction error at various flux levels when estimating a single petal mode of 150 nm using a single wavelength of 700 nm (Mono-GD) and using phase unwrapping at 600 and 700 nm. Then, 100 photon noise samples are taken at each considered number of input photons. \rev{Coloured areas correspond to $\pm1$ standard deviation.}}
    \label{fig:unwrapnoise}
\end{figure}

The increase in dynamic range provided by two-wavelength phase unwrapping will come with a hit in photon noise sensitivity due to the noise propagation properties of Equation \eqref{eq:phaseunwrap}.
This is illustrated in Figure \ref{fig:unwrapnoise}.
A single petal is displaced by 150 nm and reconstructed using a single wavelength of 700 nm and using the phase unwrapping algorithm with 600 and 700 nm.
Various photon levels were considered and for each 100 photon noise realizations were reconstructed to produce the statistics.
The variance of the phase unwrapping method is significantly greater than the monochromatic reconstruction.
There are methods proposed in the literature \rev{that can increase the dynamic range without the added increase in} noise propagation by making use of more wavelengths and/or more complicated algorithms; however, a full investigation is beyond the scope of this paper \rev{\citep{wagner_direct_2000, guo_robust_2022, costa_quantitative_2010}.}
Overall, this \rev{result} suggests that it is not useful to apply phase unwrapping when considering small errors.
In practice, we can \rev{come up with an approach where} the multi-wavelength reconstruction (as discussed in Sections \ref{dr} and \ref{section:photonnoise}) \rev{is used} for improved reconstruction under smaller errors and then switch to a phase unwrapping reconstruction in a large error regime.
In this way, the multi-wavelength information can be exploited differently, depending on the current conditions.

\section{Discussion and conclusions} \label{dandc}

 The ZWFS is a highly sensitive wavefront sensor and  a good candidate for the next generation of extreme adaptive optics (XAO) systems, but enhancing its usability in terms of dynamic range and robustness is critical.
This work uses simulations to investigate  the potential of using multi-wavelength measurements to improve the performance of the ZWFS.
A non-linear gradient descent reconstructor has been set up, with the ability to exploit the information from the different wavelengths.
Three aspects of improvement were explored.
Firstly, it was examined if the reconstructor can use the phase diversity from having multiple wavelengths and increase the dynamic range of the ZWFS.
It was found that for a scalar ZWFS, this can offer a significant gain.
For the vector ZWFS, however, no improvement was observed. This suggests that the additional diversity from multiple wavelength does not further increase the dynamic range provided by the vector concept. 
Secondly, the noise performance was investigated.
Our study shows how measuring multiple wavelengths increases the number of available photons to estimate the wavefront without compromising the reconstruction accuracy due to chromatic errors, improving the overall robustness against photon noise.
Finally, multi-wavelength phase unwrapping was implemented to reconstruct large discontinuities.
Combining a vector ZWFS, gradient descent reconstructor, and phase unwrapping algorithm enables the estimation of petal errors, but the extended dynamic range comes with a trade-off in the noise performance.

\rev{Overall, using multiple wavelengths opens up a new degree of freedom in wavefront sensing that is enabled by new algorithms, such as the accelerated gradient descent wavefront reconstructor. For  the ZWFS in particular, it appears that  combining the vector ZWFS concept with multi-wavelength measurements offers an interesting solution for the next generation of telescopes in particular, since it makes the best use of all the advantages considered here. Other future optimizations are still possible, such as the design of the mask and the cost function.}

\rev{The proposed multi-wavelength wavefront sensing scheme can be implemented with existing technologies, such as dichroics and IFUs. Nevertheless, new technologies such as MKIDs, offer a more optimal implementation. Because of its various advantages, such as inherent wavelength resolution and noise performance, a full test bed is currently being setup at SRON, combining multiple wavelengths in the visible, a deformable mirror, and a ZWFS together with an MKID array to demonstrate the potential for multi-wavelength wavefront sensing and set up a prototype for the  methods investigated here.}

\section*{Data availability}
The source code used in this work is available at \url{https://github.com/darcism/multiwavelength-zernike}.

\begin{acknowledgements}
This work is financially supported by the Netherlands Organisation for Scientific Research NWO (Vidi 213.149). \rev{We thank the referee for their time and effort, and for their comments which helped improve the quality of this paper.} 
\end{acknowledgements}

\bibliographystyle{aa} 
\bibliography{references}

\end{document}